\documentclass{iau}
\usepackage{graphicx}
\usepackage{natbib}
\usepackage{hyperref}

\title[ULLYSES and XShootU] 
{ULLYSES and Complementary Surveys of Massive Stars in the Magellanic Clouds}

\author[Paul A. Crowther]   
{Paul A. Crowther$^1$}

\affiliation{$^1$Department of Physics \& Astronomy, University of Sheffield\\ Hounsfield Road, Sheffield, S3 7RH, UK\\ email: {\tt paul.crowther@sheffield.ac.uk}}

\pubyear{2022}
\volume{361}  
\pagerange{??}
\jname{Massive Stars Near and Far}
\editors{N. St-Louis, J.S. Vink \& J. Mackey, eds.}
\begin{document}

\maketitle

\begin{abstract}
An overview is provided of the scientific goals of the Magellanic Cloud component of the STScI Directors Discretionary UV initiative ULLYSES, together with the complementary spectroscopic survey XShootU (VLT/Xshooter) and other ancillary datasets. Together, ULLYSES and XShootU permit the first comprehensive, homogeneous study of wind densities and velocities in metal-poor massive stars, plus UV/optical spectroscopic libraries for population synthesis models and a large number of interstellar sight-lines towards the Magellanic Clouds.
\keywords{
Stars: early-type - 
Stars: massive - 
Stars: evolution - 
Stars: winds, outflows - 
Stars: abundances - 
Stars: fundamental parameters}
\end{abstract}

\firstsection 
\section{Introduction}

The evolution of single and binary massive stars is inherently linked to a broad range of astrophysical transients involving their ultimate fate, including core collapse supernovae and gravitational wave mergers such as kilonovae. However, many details of their evolution remain unclear, including mass-loss, internal mixing, and close binary evolution. 

In the Milky Way, initiatives such as the Galactic O Star Spectroscopic Survey \citep[GOSSS][]{2019yCat..22240004M} together with extensive high-resolution UV spectroscopy from IUE/SWP \citep{1989ApJS...69..527H} provide reference datasets for the study of mass-loss at solar composition, and UV templates for metal-rich star-forming galaxies.

However, some exotic astrophysical phenomenon are preferentially observed in metal-deficient host galaxies (superluminous supernovae, long gamma-ray bursts) while typical high-redshift star-forming galaxies are metal-poor. Initiatives such as the VLT FLAMES Tarantula Survey \citep[VFTS][]{2011A&A...530A.108E} have provided high quality optical spectroscopy of large numbers of metal-poor massive stars, but extensive UV spectroscopy of individual stars has remained elusive until now.

In this review I will provide an overview of a new UV initiative, ULLYSES, administered under the STScI Director's Discretionary scheme, which seeks to address this deficiency, focusing on Magellanic Clouds targets. I will also summarise complementary surveys such as XShootU which seek to exploit the full potential of the programme.

\begin{figure}[ht]
\begin{center}
\includegraphics[width=0.7\textwidth,angle=-90, bb=50 80 540 770]{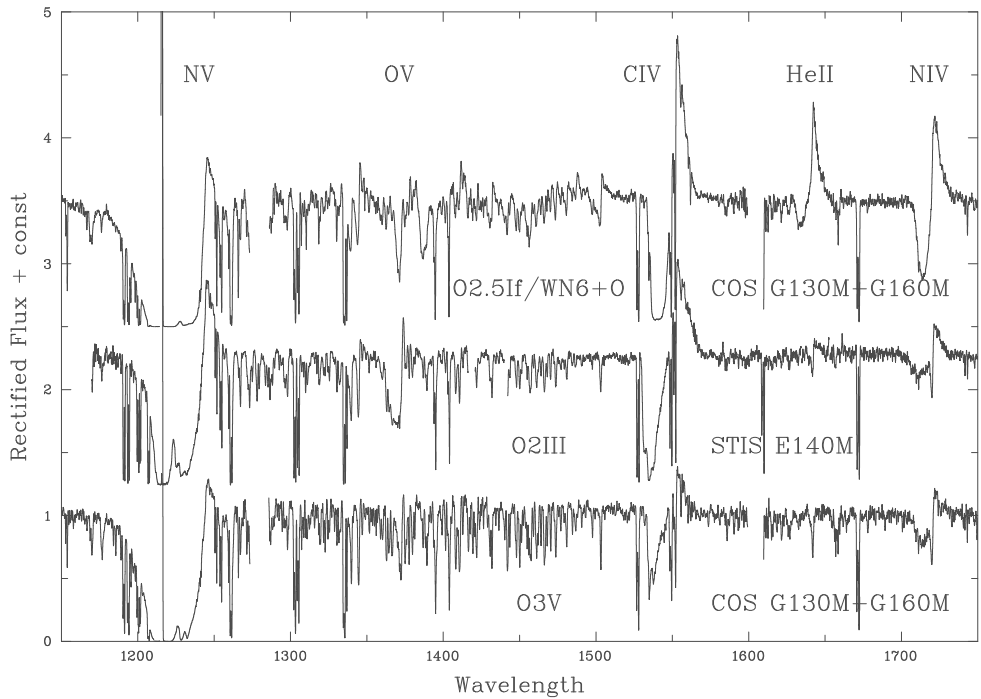}
\caption{Representative HST UV spectra of LMC early O stars (VFTS 482 = Mk~39; Sk$-67^{\circ}$ 211 = HDE 269810; PGMW 3058 from top to bottom) from ULLYSES. These spectra also highlight the forest of Fe\,{\sc v-vi} lines in the $\lambda\lambda$1250--1500 region for such stars.}
\label{earlyO}
\end{center}
\end{figure}

\section{UV Legacy stellar physics programme}

In 2018 the STScI Director Ken Sembach convened a science definition Working Group, chaired by Sally Oey, to identify a UV Legacy programme on star formation and associated stellar physics, recognising the finite lifetime of HST's large aperture UV spectroscopic capability. 

\begin{figure}[ht]
\begin{center}
\includegraphics[width=\textwidth]{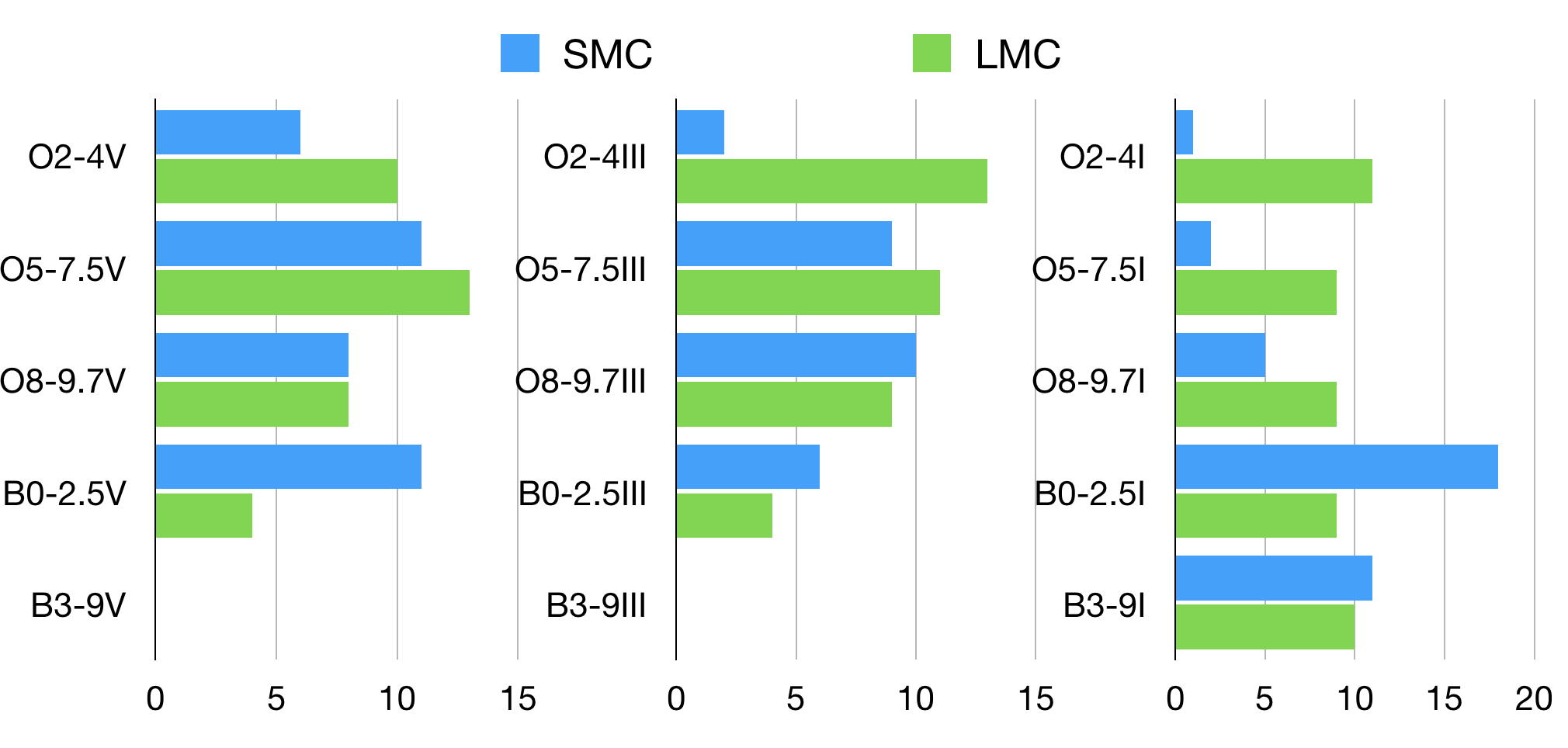}
\caption{Subtype distribution of SMC (blue/dark grey) and LMC (yellow/light grey) OB stars from ULLYSES. A few Wolf-Rayet stars, Of/WN stars and related objects are also included in the survey. ULLYSES+ targets are excluded from these charts, some of which have uncertain luminosity class.}
\label{MC-SpT}
\end{center}
\end{figure}

The Hubble UV Legacy Library of Young Stars as Essential Standards (ULLYSES) programme is based on the Working Group's recommendations, and is devoting $\sim$1000 HST orbits to an ultraviolet spectroscopic library of young stars, with Julia Roman-Duval responsible for leading the core implementation team, supported by a scientific advisory committee. This overview focuses on half of the programme involving the acquisition of high quality, high resolution COS and STIS observations of massive stars in metal-poor environments, predominantly in the Large and Small Magellanic Clouds, with present-day metallicities of 1/2 $Z_{\odot}$ and 1/5 $Z_{\odot}$, respectively.

Amongst nearby metal-poor star-forming galaxies, only the Large and Small Magellanic Clouds with distances of 50 kpc \citep{2013Natur.495...76P} and 62 kpc \citep{2014ApJ...780...59G}, respectively, are sufficiently close for medium to high resolution UV spectroscopy with HST.

\subsection{ULLYSES science}

The total ULLYSES sample comprises $\sim$250 massive stars in the Magellanic Clouds, primarily with O and B spectral types, plus a few Wolf-Rayet stars and related objects
\cite{2020RNAAS...4..205R}. A small subset of targets are known binary systems, \citep[e.g. Sk$-67^{\circ} 105$,][]{2003MNRAS.338..141O}, while others have been identified as binaries since the sample was finalised \citep[e.g. VFTS 66,][]{2020A&A...634A.118M}.
All targets include the 'standard' far ultraviolet region (Ly$\alpha$ to N\,{\sc iv} $\lambda$1718), for which representative spectra are presented in Fig.~\ref{earlyO} highlighting stellar wind (N\,{\sc v} $\lambda\lambda$1238--42, C\,{\sc iv} $\lambda\lambda$1548--51), photospheric iron forest and interstellar features. The diagnostic-rich $\lambda\lambda$900--1200 window is included for a subset of the sample (FUSE or COS/G130M/1096) while the STIS/E230M grating is added for late O and B supergiants (1978 setting, plus 2707 setting for late B supergiants).

As of June 2022, Data Release 5 (DR5) of ULLYSES includes HST spectroscopy of 233 massive stars in the Magellanic Clouds, plus FUSE spectroscopy of 122 targets. The spectral type distribution of SMC and LMC OB stars from the complete ULLYSES survey are presented in Fig.~\ref{MC-SpT}. DR5 also includes archival UV datasets for stars which were not included in the original target list (S/N or spectral coverage or uncertain luminosity class). I shall refer to these as ULLYSES+ targets throughout this overview.

UV spectroscopy provides unique information on the wind properties of early-type stars, specifically wind clumping \citep[e.g.][]{2022arXiv220211080B}. Theoretical studies \citep{2001A&A...369..574V} point to reduced wind densities in metal-poor environments, since early-type winds are radiatively-driven \citep{2021arXiv210908164V}. However, empirical studies spanning the Milky Way, LMC and SMC \citep[e.g.][]{2007A&A...473..603M} have been largely based on $H\alpha$ which is difficult to determine for stars with relatively weak winds and is insensitive to clumping. 

\begin{figure}[ht]
\begin{center}
\includegraphics[width=\textwidth]{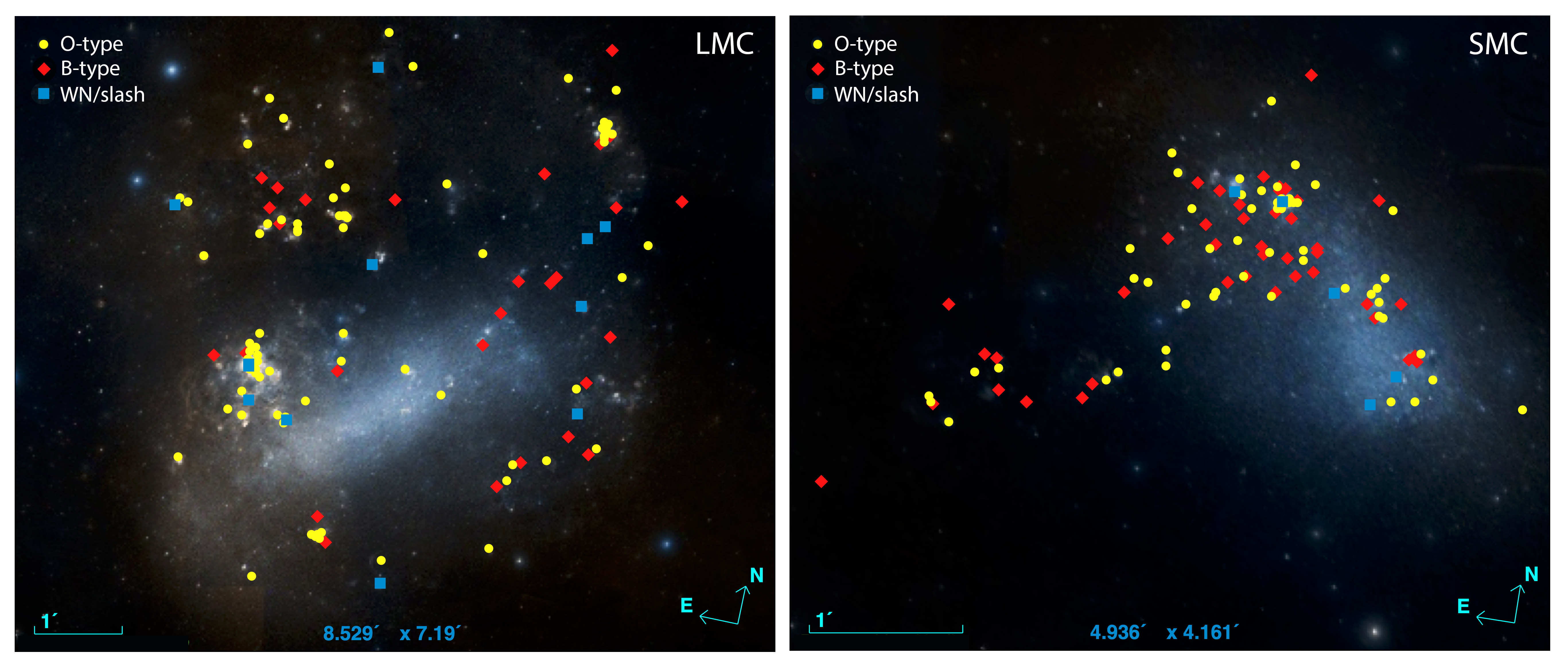}
\caption{Locations of the ULLYSES targets in the LMC (left) and SMC (right). Circles are O-type stars, diamonds are B-type stars, and squares are Wolf-Rayet and Of/WN stars. The background consists of DSS2 colour images. Scales of 8.5$^{\circ}\times 7.2^{\circ}$ and 4.9$^{\circ}\times4.2^{\circ}$ for the LMC and SMC correspond to physical scales of 7.4$\times$6.5 kpc and 5.3$\times$4.6 kpc, given their distances of 50 kpc and 62 kpc, respectively.}
\label{locations}
\end{center}
\end{figure}

ULLYSES offers the prospect of a definitive determination of stellar wind properties of high mass stars at low metallicities, with the far-UV region especially important since it includes unsaturated P Cygni profiles such as P\,{\sc v} $\lambda\lambda$1118-1128 \citep{2002ApJ...579..774C}.

The primary diagnostic for wind velocities in early-type stars involves the blueward trough of UV P Cygni profiles (e.g. C\,{\sc iv} $\lambda\lambda$1548--51) which ULLYSES/ULLYSES+ permit the extension of Galactic calibrations \citep{1990ApJ...361..607P} to metal poor environments (C. Hawcroft et al. in prep), which have historically involved theoretical predictions \citep{1992ApJ...401..596L}. Comparisons of wind properties between the Milky Way, LMC and SMC \citep[e.g.][]{2007A&A...473..603M} often include (reduced) wind momenta, $\dot{M} v_{\infty} R^{0.5}$ \citep{1999A&A...350..970K}, yet commonly adopt historical Milky Way OB scalings \citep{1995ApJ...455..269L} which may not apply to metal-poor counterparts \citep{1998MNRAS.300..828P}.

Together ULLYSES/ULLYSES+ comprise $\sim$15\% of the total O star population in the SMC \citep{2018A&A...618A..17K}, versus 4\% of the LMC O star population, whose star-formation rate is a factor of five times higher \citep{2008ApJS..178..247K}. The scarcity of SMC early O-type (super)giants in Fig.~\ref{MC-SpT} arises from its modest star formation rate \citep[see also][]{2021A&A...646A.106S}. 

ULLYSES promises to provide important diagnostics beyond mass-loss through abundance determinations of CNO elements. To date, optical studies of O stars to date have permitted the determination of He and N abundances \citep{2008ApJ...676L..29H}, with the addition of far-UV spectroscopy critical to the determination of C/N and O/N abundances, which will help to constrain mixing in metal-poor massive stars.

Beyond stellar astrophysics, ULLYSES will also significantly extend the number of interstellar Magellanic Cloud H\,{\sc i} and metal sight-lines beyond those from previous surveys \citep[e.g. METAL,][]{2019ApJ...871..151R}, plus H$_{2}$ column densities \citep{2002ApJ...566..857T, 2012ApJ...745..173W}. The spatial location of the ULLYSES Magellanic Cloud targets is presented in Figure~\ref{locations}. The latter Lyman-Werner bands contaminate far-UV spectral diagnostics to varying degrees \citep{2002ApJS..141..443W}.

\begin{figure}[ht]
\begin{center}
\includegraphics[width=0.8\textwidth]{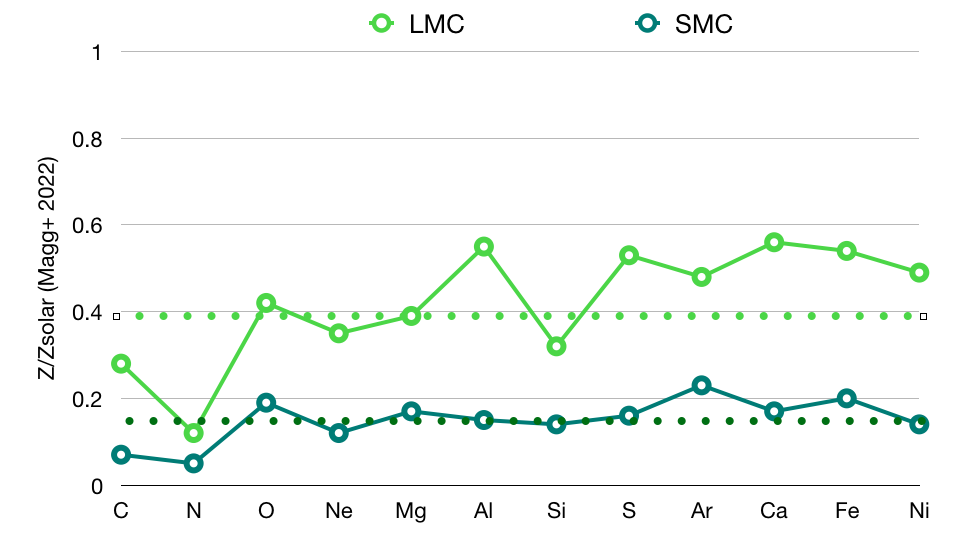}
\caption{Present day abundances of the LMC (pale green/grey) and SMC (dark green/grey) from H\,{\sc ii} regions, stars and supernova remnants (see J. Vink et al. in prep for details) with respect to the solar value \citet{2022A&A...661A.140M}.}
\label{MC-metallicity}
\end{center}
\end{figure}

\subsection{UV population synthesis at low metallicity}

Historically, UV population synthesis tools \citep[e.g. Starburst99,][]{2014ApJS..212...14L} have incorporated templates of Milky Way OB stars from extensive samples from IUE/SWP \citep{1989ApJS...69..527H} but Magellanic Cloud OB samples from HST have been so scarce \citep{2000PASP..112.1243W} that combined LMC/SMC templates have been adopted, or theoretical spectra have been used \citep{2004ApJ...615...98R}. Large numbers of ULLYSES targets permits the construction of templates at LMC and SMC metallicities, of application to metal-poor galaxies \citep[e.g. CLASSY,][]{2022arXiv220307357B}.

In general, scaled Solar abundances are adopted for stars in the Magellanic Clouds, but there is considerable element-to-element variation from H\,{\sc ii} regions, stars and supernova remnants (for details see J. Vink et al, in prep), as shown in Figure~\ref{MC-metallicity}. Adopting the latest \cite{2022A&A...661A.140M} solar values, the average present day LMC metallicity is 0.4 $Z_{\odot}$, with [O/H]=$-$0.4 dex and [Fe/H]=$-$0.3 dex, while the average SMC metallicity is 0.15 $Z_{\odot}$ with [O/H]=[Fe/H]=$-$0.7 dex.

\begin{figure}[ht]
\begin{center}
\includegraphics[width=0.45\textwidth,angle=-90,bb=40 50 540 760]{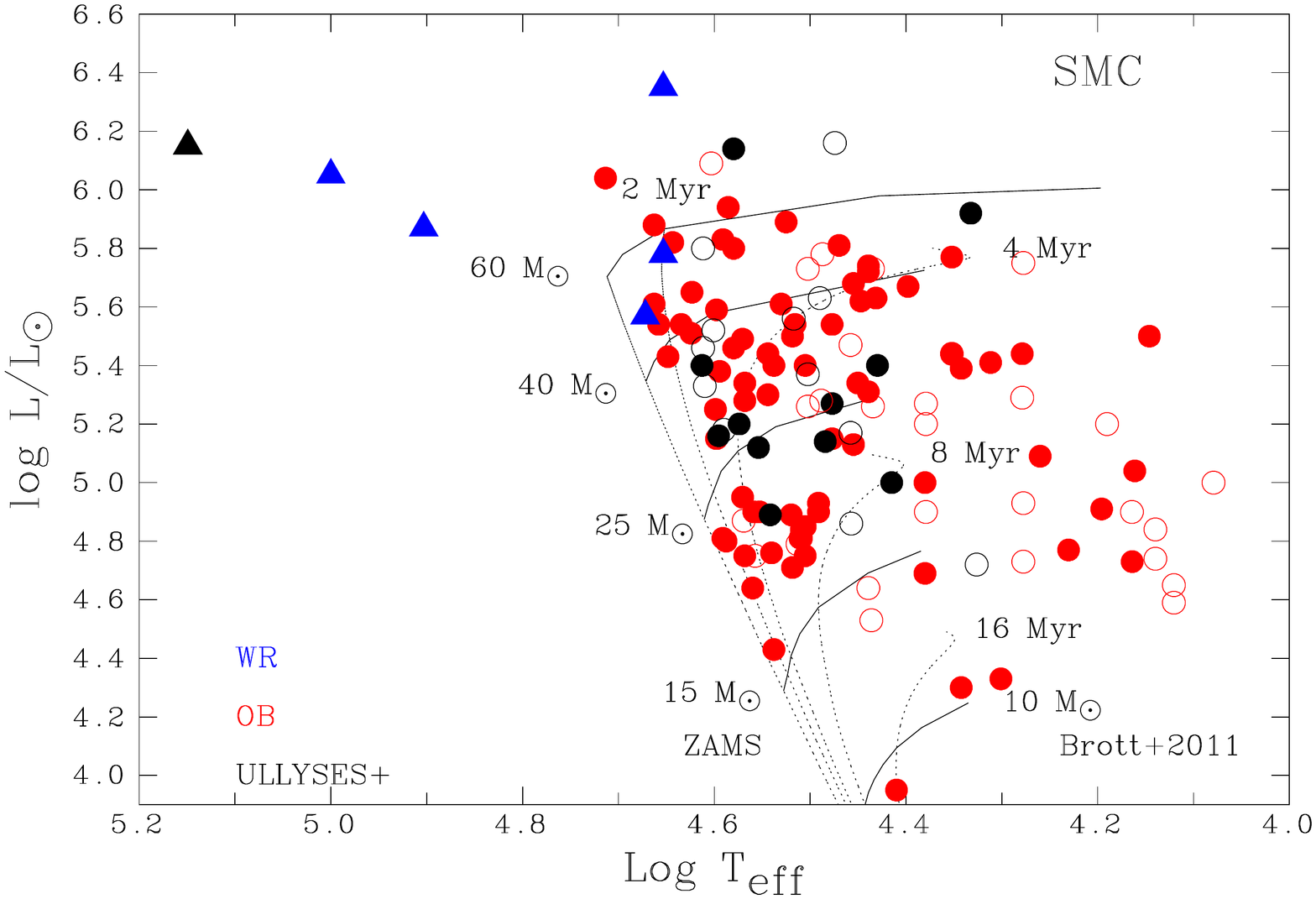}
\includegraphics[width=0.45\textwidth,angle=-90,bb=40 50 540 760]{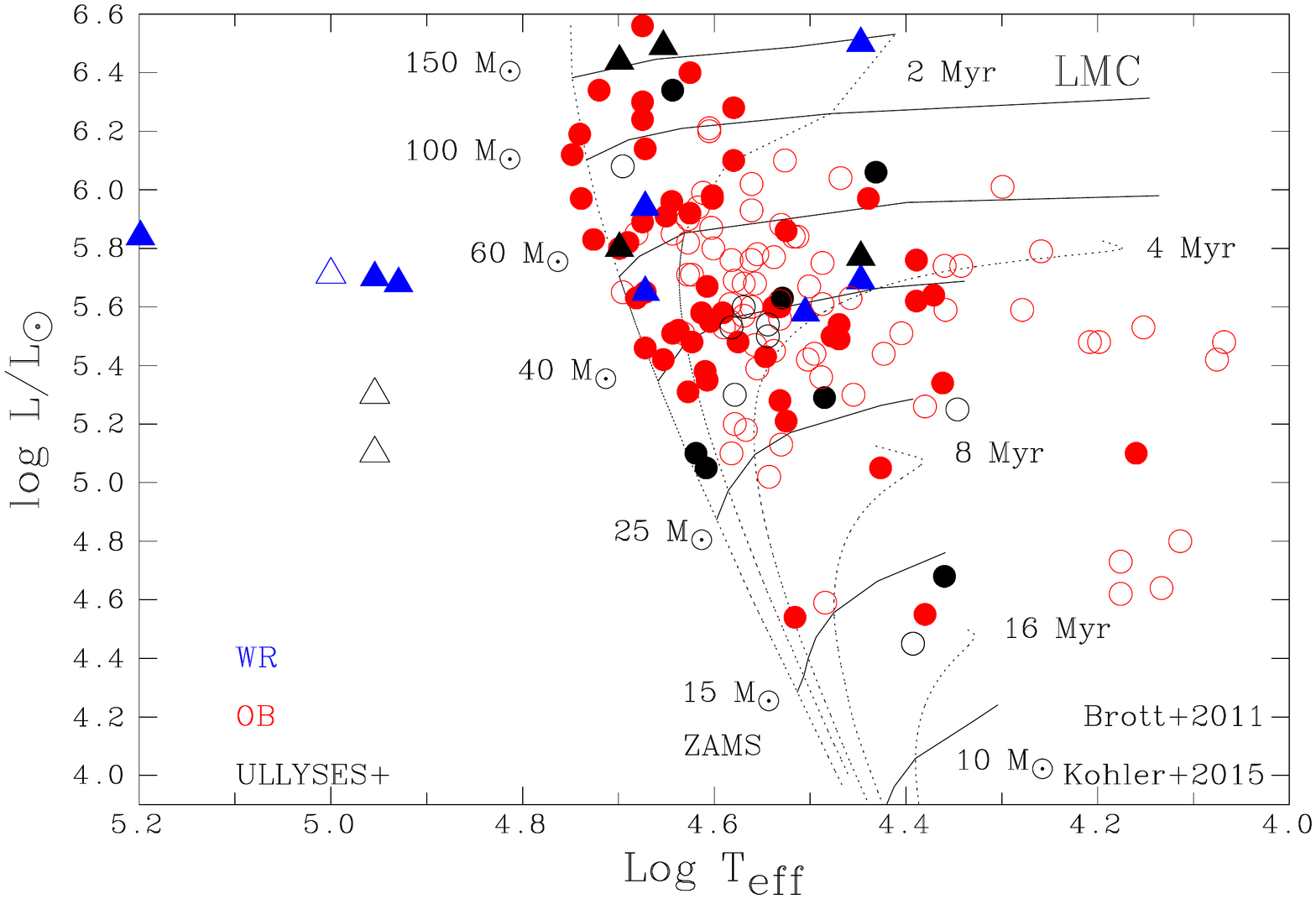}
\caption{Hertzsprung-Russell diagram of SMC (top) and LMC (bottom) ULLYSES targets, OB stars in circles (red), WR stars in triangles, based on previous spectroscopic fitting (filled symbols) or calibrations (open symbols), plus ULLYSES+ stars in black. $0.2 Z_{\odot}$ or $0.5 Z_{\odot}$ evolutionary tracks and isochrones are from \citet{2011A&A...530A.115B} and \citet{2015A&A...573A..71K} for very massive stars in LMC.}
\label{HRD}
\end{center}
\end{figure}

\section{Complementary datasets}

Extensive optical and IR photometry exists for bright stars in the Magellanic Clouds, including Spitzer IRAC+MIPS \citep[e.g.][]{2009AJ....138.1003B}, and the VISTA VMS near-IR public survey \citep{2011Msngr.144...25C}. However, complementary optical spectroscopic datasets are required to exploit the full potential of ULLYSES.

Optical spectroscopic studies for the majority of ULLYSES targets have been undertaken \citep[e.g][]{2009ApJ...692..618M, 2011A&A...530A.108E}, albeit involving heterogeneous tools and datasets, while properties for the remainder can be obtained from contemporary spectral type calibrations \citep{2013A&A...558A.134D}. In Figure~\ref{HRD} we present Hertzsprung-Russell diagrams for ULLYSES(+) targets in each galaxy, together with evolutionary tracks and isochrones from \citet{2011A&A...530A.115B} and \citet{2015A&A...573A..71K}. The typical mass range for SMC OB stars is 15--60 $M_{\odot}$ versus 25--150 $M_{\odot}$ for LMC OB stars.

\subsection{XShootU}

Consequently, the large Hubble ULLYSES award has motivated the acquisition of complementary homogeneous spectroscopy, in particular the XShootU initiative, coordinated by the IAU Commission G2 (chaired by Jorick Vink between 2018--2021). XShootU involves VLT/XShooter spectroscopy of all Magellanic Cloud ULLYSES targets. Further details of XShootU are provided in J. Vink et al (in prep).

To date, calibrated reductions for $\sim$70\% of optical (UBV + VIS arms) XShooter datasets have been provided to the XShootU consortium by Working Group 2 (coordinator Hugues Sana) via the March 2022 eDR1 release. The remaining 30\% of optical datasets and NIR arm will follow in due course.

\begin{figure}[ht]
\begin{center}
\includegraphics[width=\textwidth]{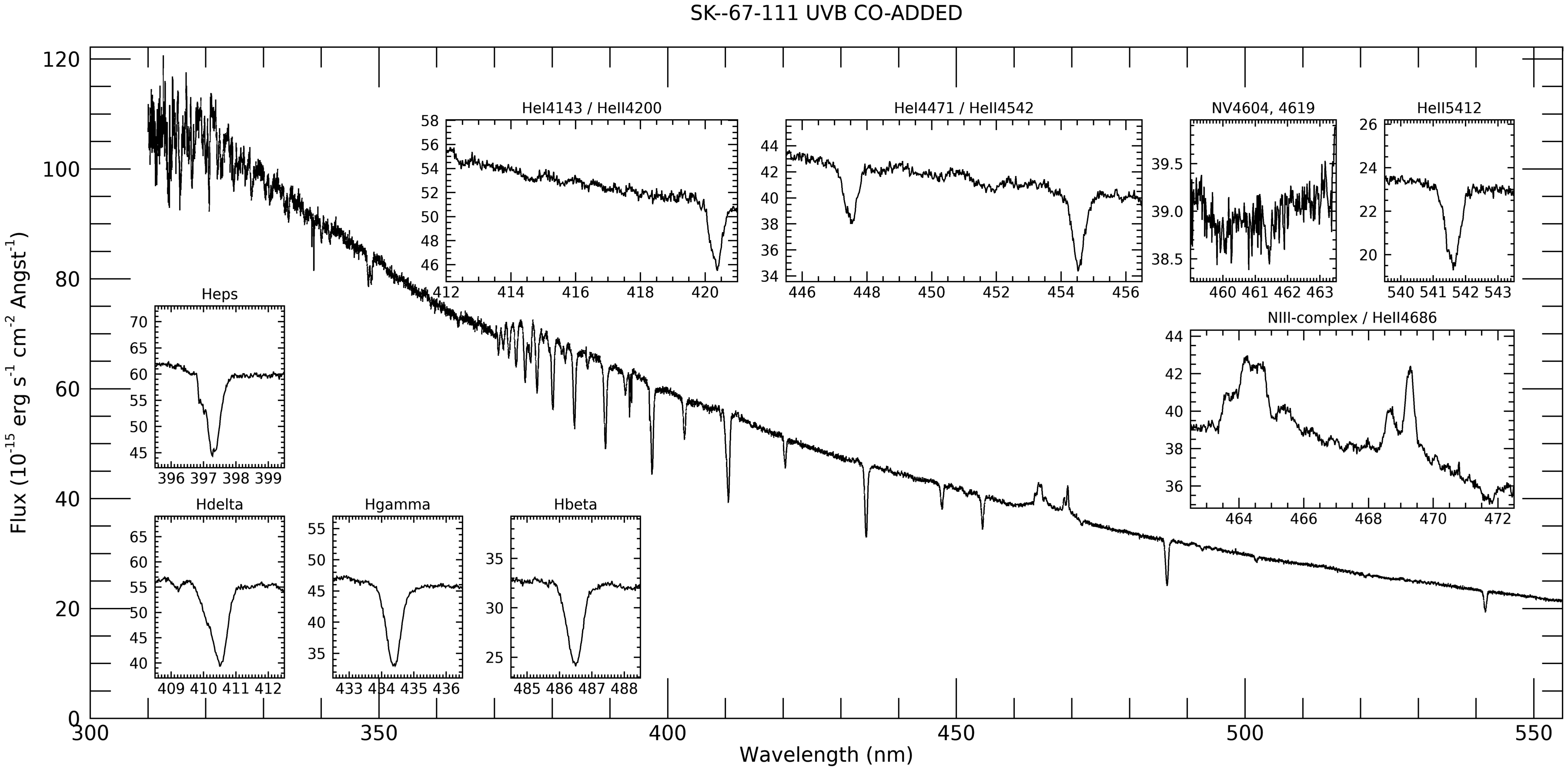}
\caption{VLT/Xshooter UVB spectroscopy of Sk$-67^{\circ}$ 111 (O6\,Iaf) from the XShootU initiative}
\label{sk67_111}
\end{center}
\end{figure}

A representative Xshooter UBV observation of Sk$-67^{\circ}$111 (O6\,Iaf) in the LMC is presented in Fig.~\ref{sk67_111}, illustrating the high data quality of XShootU, which will permit the uniform determination of stellar temperatures, surface gravities, helium abundances and rotational velocities for the entire ULLYSES sample (ULLYSES+ targets were not acquired).

\begin{figure}[ht]
\begin{center}
\includegraphics[width=0.49\textwidth]{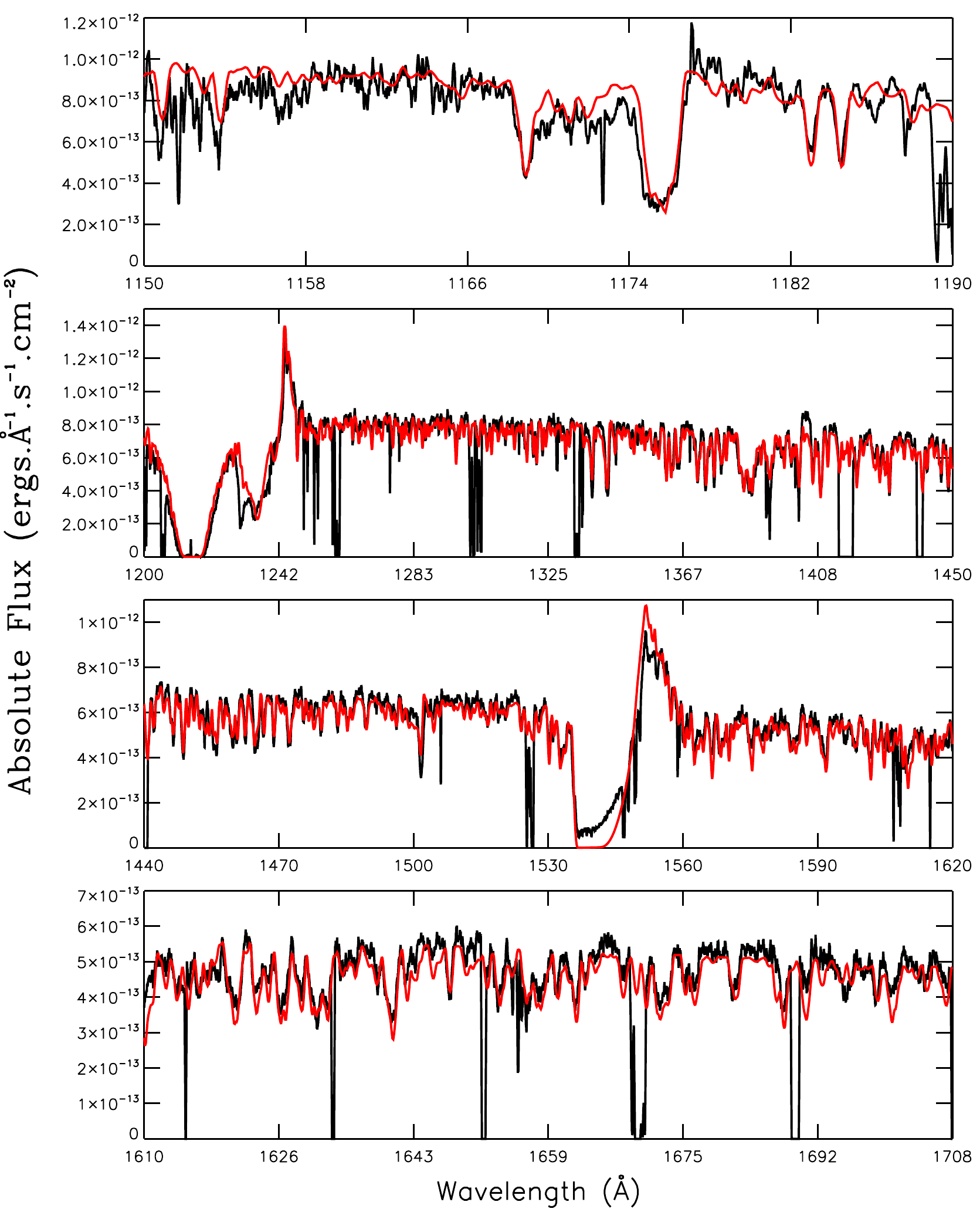}
\includegraphics[width=0.49\textwidth]{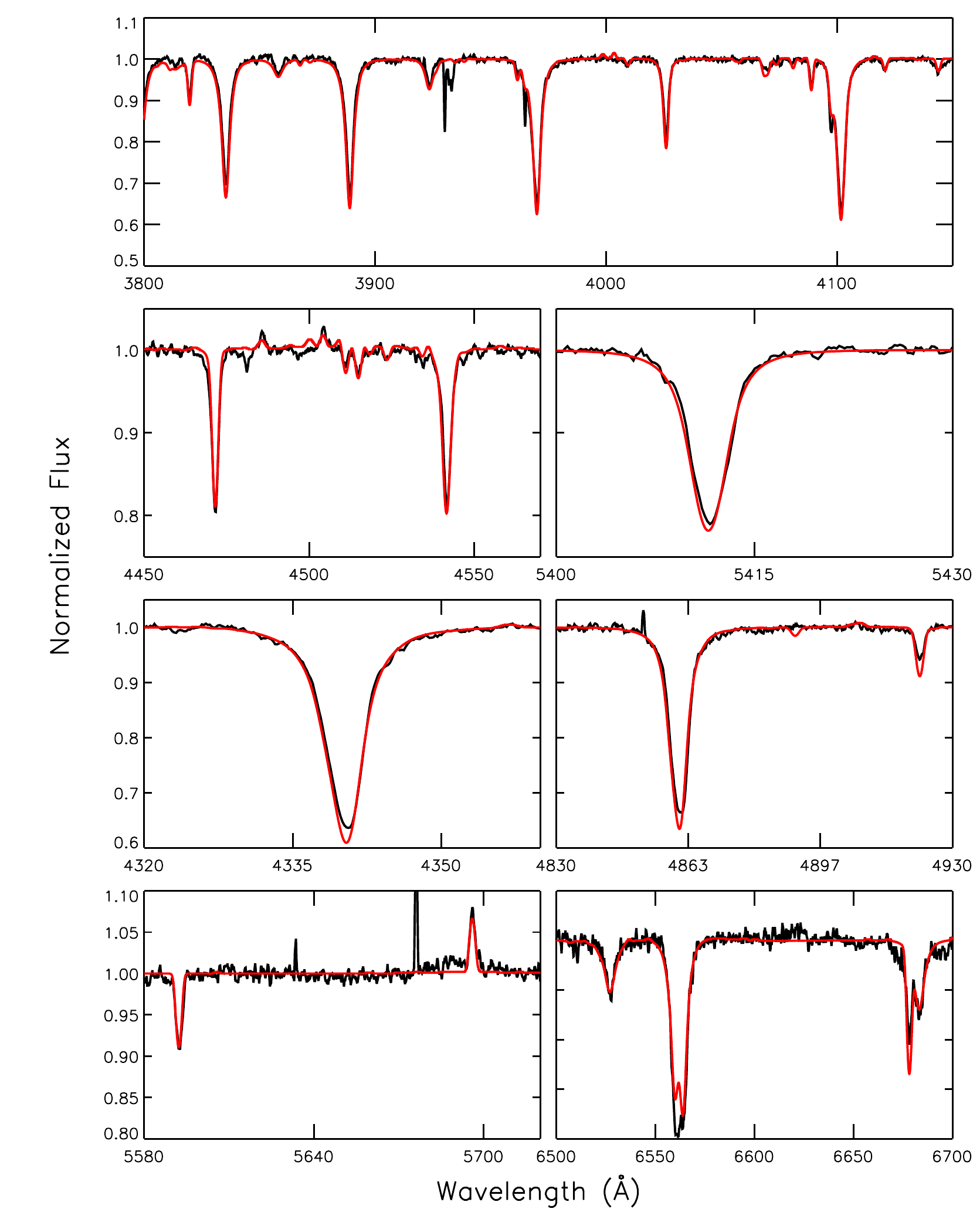}
\caption{Preliminary CMFGEN fits (red/grey) to UV (HST/STIS/E140M from ULLYSES, left) and optical (VLT/Xshooter from XShootU, right) spectroscopy (black) of the LMC star Sk$-68^{\circ}$ 16 (O7\,III), courtesy of JC Bouret.}
\label{sk68_16}
\end{center}
\end{figure}

Together, ULLYSES and XShootU offer the potential to characterize the physical and wind properties of an unprecedented sample of Magellanic Cloud early-type stars. Preliminary CMFGEN \citep{1998ApJ...496..407H} spectral fits to Sk$-68^{\circ}$ 16 (O7\,III) are presented in Fig.~\ref{sk68_16}, courtesy of JC Bouret, although FASTWIND \citep{2020A&A...642A.172P} and PoWR \citep{2002A&A...387..244G} are also well suited to spectroscopic analysis of early-type stars with winds.

UV spectroscopic studies of high ionization P Cygni resonance lines (e.g. O\,{\sc vi} $\lambda\lambda$1031, 1037) N\,{\sc v} $\lambda\lambda$1238, 1242) are also sensitive to X-rays \citep[e.g.][]{1997A&A...321..531T}, which are known to originate in the winds of early-type stars \citep{2022arXiv220316842R}. To date the majority of X-ray studies of OB stars have involved Milky Way stars, although deep {\it Chandra} X-ray observations of a few star-forming regions in the Magellanic Clouds have been obtained.

\begin{figure}[ht]
\begin{center}
\includegraphics[width=0.6\textwidth,angle=-90,bb=50 100 540 730]{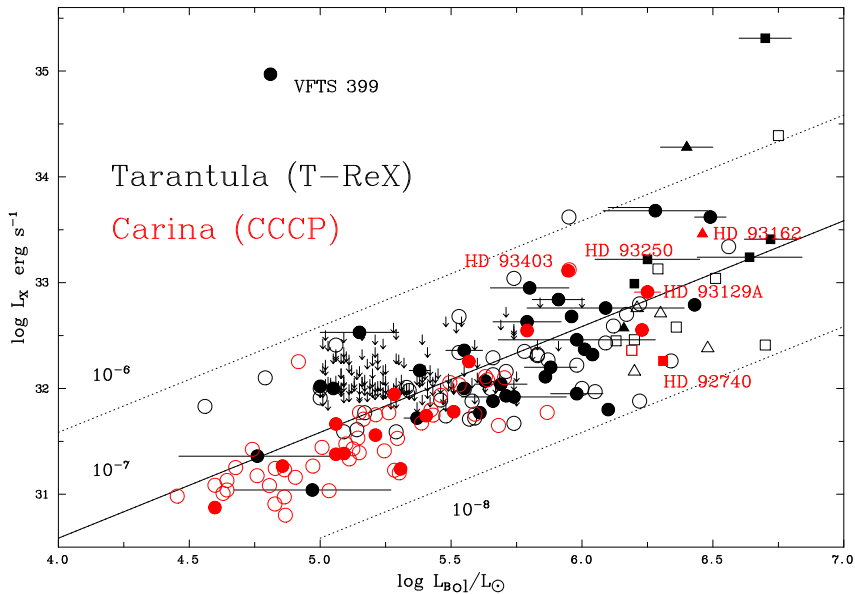}
\caption{Comparison between X-ray and bolometric luminosities for single (open) and binary (filled) O stars (circles), Of/WN stars (triangles) and WN stars (squares) in the Tarantula Nebula from XSPEC fits to T-ReX observations \citep[black]{Crowther2022} with those in the Carina Nebula from CCCP \citep[red/grey]{2011ApJS..194....7N}.}
\label{Carina}
\end{center}
\end{figure}

In particular, T-ReX (PI Leisa Townsley), has involved a 2\,Ms {\it Chandra} ACIS survey of the Tarantula Nebula, revealing over 100 detections of early-type sources. Fig.~\ref{Carina} compares bolometric to X-ray luminosities of O, Of/WN and WN stars in the Carina Nebula \citep{2011ApJS..194....7N} to the Tarantula Nebula \citep{Crowther2022} revealing similar X-ray properties for luminous stars between both environments, such that similar X-ray assumptions are recommended for metal-poor massive stars.

\begin{figure}[ht]
\begin{center}
\includegraphics[width=0.8\textwidth]{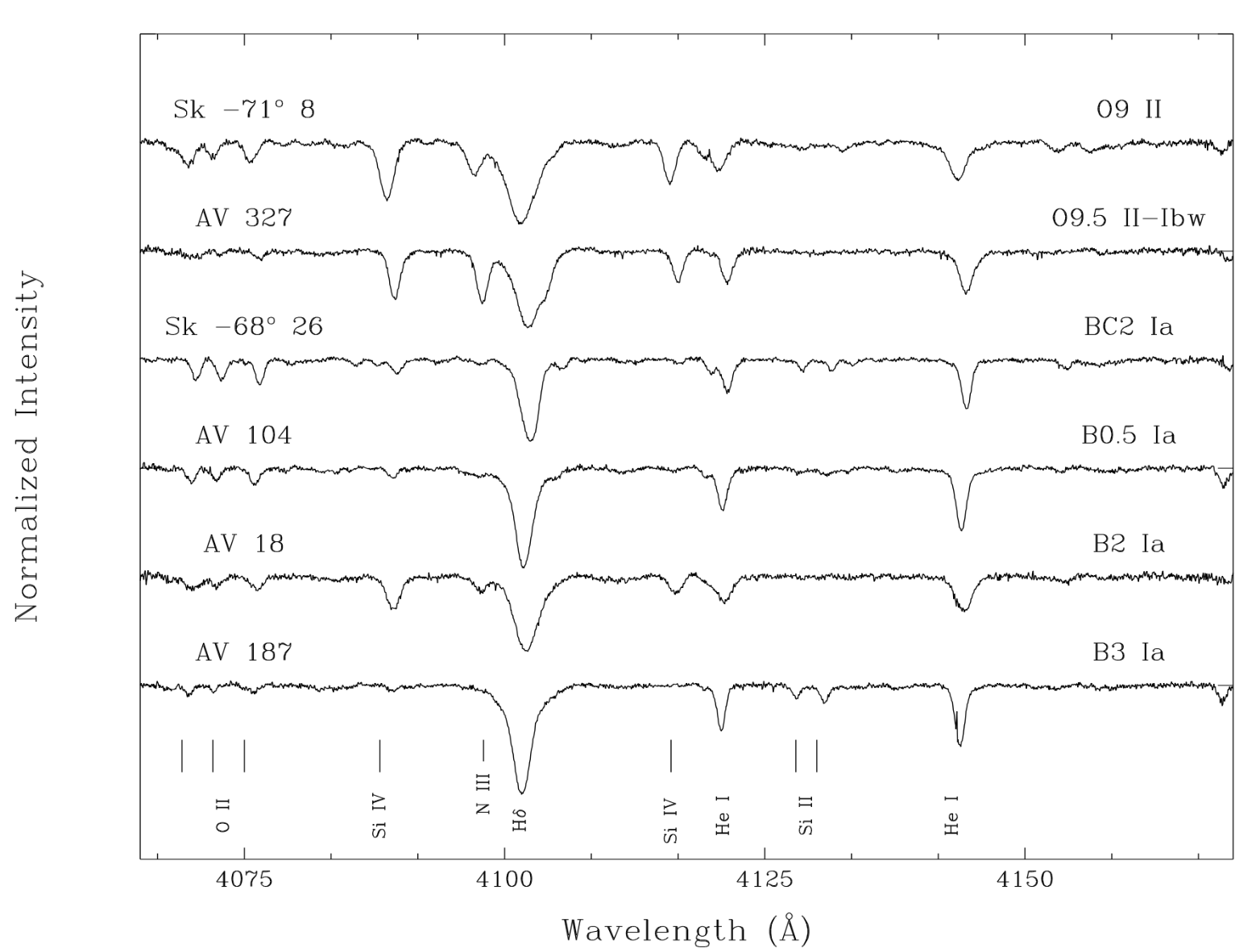}
\caption{Representive Magellan/MIKE spectroscopy of slowly rotating LMC and SMC ULLYSES targets from Dec 2021, courtesy Nidia Morrell.}
\label{MIKE}
\end{center}
\end{figure}

\subsection{Magellan/MIKE}

The spectral resolution of Xshooter (R$\sim$7,000) is sufficient for the majority of ULLYSES targets, although slowly rotating OB stars would benefit from higher spectral resolution in order to resolve spectral features and characterize their rotation rates. Nidia Morrell has obtained complementary Magellan/MIKE spectroscopy (R$\sim$35,000--40,000) of 23 slowly rotating ULLYSES targets in Dec 2021, with an additional 3 night allocation for Dec 2022. Representative LMC and SMC examples are presented in Fig.~\ref{MIKE}.

\subsection{Population synthesis empirical templates at low metallicity}

The XShootU datasets have application beyond individual spectroscopic analysis. Population synthesis tools have historically included either theoretical spectra from model atmospheres, such as ATLAS \citep{2003IAUS..210P.A20C} or MARCS \citep{2008A&A...486..951G} or empirical spectra spanning a range of spectral coverage and resolution \citep[e.g.][]{1984ApJS...56..257J, 2003A&A...402..433L}. 

In recent years, attempts have been made to extend templates to metal-poor populations across a range of temperatures and gravities \citep[e.g. MaStar,][]{2020MNRAS.496.2962M}. These have been successful for old, low mass populations within the Milky Way \citep{2022MNRAS.509.4308H} but lack metal-poor templates amongst massive stars, which XShootU will address.

\section{Summary and outlook}

We provide a brief overview of the Magellanic Cloud component of the UV Legacy programme ULLYSES and complementary optical spectroscopic datasets, notably the VLT/Xshooter survey XShootU. Together, these offer the prospect of characterising the physical, wind and chemical properties of large samples of OB stars in the LMC and SMC, together with providing UV and optical spectroscopic templates for unresolved metal-poor populations and extensive interstellar sight-lines at intermediate to high spectral resolution.

In the near future, VISTA/4MOST will provide very large numbers of optical spectroscopic datasets for early-type stars in the Magellanic
Clouds via the 1001MC survey \citep{2019Msngr.175...54C} although the prospect for high resolution UV spectroscopy of individual stars at metallicities below the Magellanic Clouds requires a much larger aperture space telescope such as the Large UV/Optical/IR Surveyor (LUVOIR).

\begin{acknowledgements}
Thanks to the following for their major contributions to ULLYSES and XshootU: Ken Sembach (STScI Director), Sally Oey (Working Group chair), Julia Roman-Duval (ULLYSES implementation team lead), Jorick Vink (IAU G2 Commission chair), plus JC Bouret for preliminary fits to Sk $-68^{\circ}$ 16 and Nidia Morrell for proposing and observing the Magellan MIKE datasets. PAC is supported by the Science and Technology Facilities Council research grant ST/V000853/1 (PI. V. Dhillon).
\end{acknowledgements}

\bibliographystyle{apj}
\bibliography{refs}




\begin{discussion}
\discuss{Leitherer}{Paul, you mentioned abundances (CNO) but how far away are we from determining iron abundances in the ultraviolet?}

\discuss{Crowther}{Although fits to the UV forest in Magellanic Cloud stars for an adopted iron abundance are pretty good, accurate abundances from ULLYSES are very challenging given the complexity of atomic data involved, so alternative methods are preferred (e.g. optical Fe\,{\sc iii} lines in B stars, Thompson et al. 2008 MNRAS 383, 729).}

\discuss{XXX}{Does the ULLYSES target list include stars which allow us to tackle the weak wind problem?}

\discuss{Crowther}{Yes, both galaxies include late O and early B dwarfs (Fig.~\ref{MC-SpT}) which have been shown to exhibit significantly weaker winds than predicted by radiatively driven wind theory, based on UV/H$\alpha$ diagnostics.}

\end{discussion}

\end{document}